\documentclass[traditabstract]{aa} 
\usepackage{graphicx}
\usepackage{txfonts}
\usepackage{natbib}
\usepackage{textcomp}
\usepackage{longtable}
\usepackage{booktabs}
\usepackage{lscape}
\usepackage{xspace}
\usepackage{tabularx}

\begin{document}
\title{Photometric AGN reverberation mapping of 3C120}

	\author{
          Francisco Pozo Nu\~{n}ez 
          \inst{1,2} 
          \and
          Michael Ramolla
          \inst{1}
          \and
          Christian Westhues
          \inst{1}
          \and
          Christoph Bruckmann
          \inst{1}
          \and 
          Martin Haas
          \inst{1}
          \and 
          Rolf Chini
          \inst{1,2}
          \and
          Katrien Steenbrugge
          \inst{2,3}		
          \and
          Miguel Murphy
          \inst{2}}
	\institute{
          Astronomisches Institut, Ruhr--Universit\"{a}t Bochum,
	  Universit\"{a}tsstra{\ss}e 150, 44801 Bochum, Germany
	  \and
          Instituto de Astronomia, Universidad Cat\'{o}lica del
          Norte, Avenida Angamos 0610, Casilla
          1280 Antofagasta, Chile
	  \and
          Department of Physics, 
          University of Oxford, 
          Keble Road,
          Oxford OX1 3RH, UK\\}


	\date{Received February 24, 2012 ; accepted July 30, 2012}
        
	\abstract{ We present the results of a five month monitoring campaign
          of the local active galactic nuclei (AGN) \object{3C120}. Observations with a median sampling of
          two days were conducted with the robotic 15cm telescope
          VYSOS-6 located near Cerro Armazones in Chile. 
          Broad band ($B$,$V$) and narrow
          band (NB) filters were used in order to measure
          fluxes of the AGN and the H$\beta$ broad line region (BLR) emission line.
          The NB flux is constituted by about
          50\% continuum and 50\% H$\beta$ emission line. To
          disentangle line and continuum flux, 
          a synthetic H$\beta$ light curve was created by
          subtracting a scaled $V$-band light curve from the NB
          light curve. Here we show that the H$\beta$ emission line
          responds to continuum variations with a rest frame 
          lag of $23.6 \pm 1.69$
          days. 
          We estimate a virial mass of the central black
          hole $M_{\rm BH} = 57 \pm 27 \cdot 10^{6}\, M_{\odot}$, by
          combining the obtained lag with the velocity dispersion of a
          single contemporaneous spectrum. 
          Using
          the flux variation gradient (FVG) method, we determined 
          the host galaxy
          subtracted rest frame 5100\AA~ luminosity at the time of our
          monitoring campaign with an uncertainty
          of 10\% 
          ($L_{\rm{AGN}} = (6.94 \pm 0.71)\times 10^{43}erg~s^{-1}$). 
            Compared with recent spectroscopic reverberation results, 
            \object{3C120} shifts in the $R_{BLR}$ - $L_{AGN}$ diagram 
            remarkably close to the theoretically expected 
            relation of $R \propto L^{0.5}$. 
          Our results demonstrate the performance of photometric 
          AGN reverberation mapping,
          in particular for efficiently determining the BLR size and 
          the AGN luminosity.
        }

		\maketitle
%

\section{Introduction}

Reverberation mapping (\citealt{1982ApJ...255..419B}), where
spectroscopic monitoring is used to measure the response delay $\tau$ of
the broad emission lines to nuclear continuum variations, has proven to
be a powerful tool to measure the average distance of the BLR clouds 
to the central source $R_{\rm{BLR}} = \tau \cdot c$.

The spectroscopy provides us also with a velocity dispersion
$\sigma_{\rm{v}}$ of the emitting gas. Then, adopting Keplerian
motion, one may estimate the enclosed mass, dominated by
the supermassive black hole (\citealt{2004ApJ...613..682P} and
references therein).

From theoretical considerations (\citealt{1990agn..conf...57N}), the
relationship $R_{\rm{BLR}} \propto L_{\rm{AGN}}^{\alpha}$, between H$\beta$ BLR
size and nuclear luminosity (5100\AA), should have $\alpha = 0.5$. This has been
investigated intensively (\citealt{1991ApJ...370L..61K,2000ApJ...533..631K,2005ApJ...629...61K,2006ApJ...644..133B,2009ApJ...697..160B}), with the latest 
result being $\alpha = 0.519^{+0.063}_{-0.066}$. 

Notably a new technique for distance determination based on
dust-reverberation 
mapping has been presented (\citealt{2002ntto.conf..235Y}).
Modifying these concepts, it has recently been proposed 
that the $R_{\rm{BLR}} \propto L_{\rm{AGN}}^{\alpha}$ relationship 
can be used as an 
alternative luminosity distance indicator. 
The intrinsic luminosity of the AGN ($L_{\rm{AGN}}$) 
can be inferred from the radius of
the BLR ($R_{\rm{BLR}}$), 
resulting in the determination of the AGN distance. 
Moreover, due to the large luminosity and the extensive 
range of redshift at which the AGNs can be observed, 
the $R_{\rm{BLR}}-L_{\rm{AGN}}$ relationship offers the opportunity 
to discriminate between 
different cosmologies and to probe the dark energy 
(\citealt{2011A&A...535A..73H}; \citealt{2011ApJ...740L..49W}). 
However, these methods require that the large 
scatter of the current $R_{\rm{BLR}}-L_{\rm{AGN}}$ relation can be reduced 
significantly, i.e. by factors up to 10, and that reverberation 
mapping of large samples can be performed efficiently.

Recently, \citet{2011A&A...535A..73H} have revisited photometric
reverberation mapping. They demonstrated for Ark120 and 
PG0003+199 (Mrk335) that this method is very efficient and 
even applicable using very small telescopes.
Broad filters are used to measure the triggering continuum,
while suitable narrow band filters catch the emission line 
response.

The estimation of the host-subtracted nuclear luminosity
$L_{\rm{AGN}}$ is 
challenging as well. One may use high-resolution imaging
data and model the host galaxy profile in order to disentangle the
nuclear flux (\citealt{2009ApJ...697..160B}, using HST imaging). 
An alternative approach is provided by the flux variation gradient
method (FVG, by \citealt{1981AcA....31..293C,1997MNRAS.292..273W}). 
This method can be easily applied to monitoring data and 
does not require high spatial resolution.

\begin{table*}
\begin{center}
\caption{Characteristics of \object{3C120}}
\label{table1}
\begin{tabular}{@{}cccccccc}
\hline\hline
$\alpha$ (2000)$^{(1)}$ & $\delta$ (2000)$^{(1)}$ & z$^{(1)}$ & $D_L^{(2)}$ & H$\beta lag^{(3)}$ & $M_{BH}^{(3)}$ & $A_B^{(4)}$ & $A_V^{(4)}$ \\
      & & & (Mpc) & (days) & ($M_{\odot}$) & (mag) & (mag) \\
\hline
04:33:11.095 & +05:21:15.620 & 0.03301 & 145.0 & 38.5 $^{+21.3}_{-15.3}$ & $55.5 \pm 26.9 \times 10^{6}$ & 1.283 & 0.986 \\
\hline
\end{tabular}
\end{center}
\tablefoottext{1}{Values from NED database.}, 
\tablefoottext{2}{\citet{2009ApJ...705..199B},}
\tablefoottext{3}{\citet{2004ApJ...613..682P},}
\tablefoottext{4}{\citet{1998ApJ...500..525S}.}
\end{table*}

\object{3C120} is a nearby Seyfert 1 galaxy
known to be strongly variable in the optical, characterized by short and
long term variations with amplitudes of up to 2 mag on a timescale of 10
years (\citealt{1979SvA....23..518L,1988AJ.....95..374W}). Subsequent
studies showed amplitude variations of about 0.4 mag
(\citealt{1997MNRAS.292..273W}) and 1.5 mag
(\citealt{2010ApJ...711..461S}) on a timescale of 4 years
monitoring. Furthermore, \object{3C120} was included in a complete eight years AGNs spectroscopic monitoring campaign conducted by \citet{1998ApJ...501...82P}. Although \object{3C120} was a lower priority object (with 50 days average intervals between observations) the spectroscopic reverberation mapping results
have shown that the H$\beta$ emission line time response allows one to
determine the BLR size with an uncertainty of about 50\%
(\citealt{2004ApJ...613..682P}). Through 3C120's redshift of $z =
0.0331$, the H$\beta$ line falls into the OIII filter at 5007\AA. Consequently, it is an ideal candidate for our instrumentation,
a robotic 15 cm refractor.

Here we present new measurements of the BLR size, host-subtracted AGN
luminosity and black hole mass based on a well-sampled photometric
reverberation mapping campaign, allowing us to
revisit the position of \object{3C120} in the BLR size - luminosity
relationship.
As a lucky coincidence, Grier et al. (2012)  have carefully monitored \object{3C120}
spectroscopically one year
after our campaign allowing us to directly compare our photometric monitoring
results with their spectroscopic results.

\begin{figure}
  \centering
  \includegraphics[angle=90,width=\columnwidth]{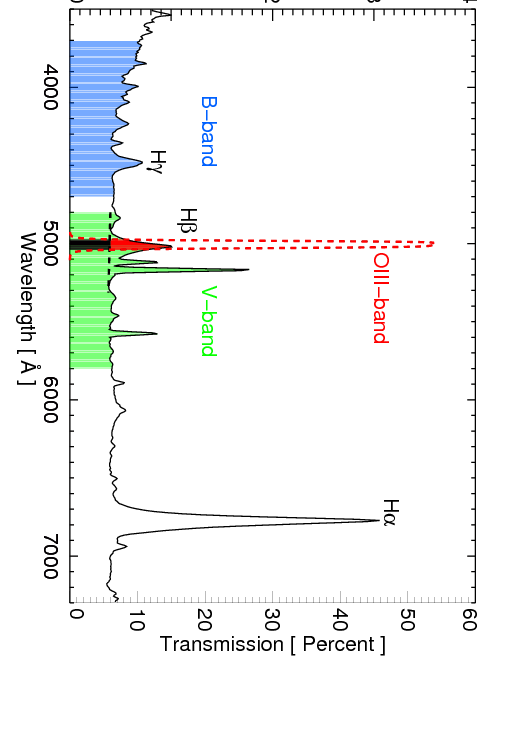}
  \caption{ 
    CAFOS spectrum of \object{3C120} obtained on Oct.$27^{th}$ 2009. 
    For illustration, the band passes of the filters used for 
    the photometric monitoring
    are shaded (blue $B$-band, green $V$-band).
    The transmission of the narrow band ([OIII] 5007\AA~filter) is indicated
    by the red dotted line. 
    While narrow band catches the redshifted H$\beta$
    line, its flux is composed of  
    about 50\%  H$\beta$ line (red shaded) and 50\% the continuum
    (black shaded).
    Note that for actual flux calculations the filter curves are
    convolved with the transmission curve of the Alta U16 CCD camera.
  }
  \label{spectrum_nbfilter}
\end{figure}
 
\section{Observations and data reduction}

The photometric monitoring campaign was conducted between October 2009
and March 2010 using the robotic 15\,cm VYSOS-6 telescope of the
Universit\"atssternwarte Bochum,
located near Cerro Armazones in Chile\footnote{More information about the
telescope and the instrument has been published by
\citet{2011A&A...535A..73H}.}

The images were reduced using IRAF\footnote{IRAF is distributed by the National Optical Astronomy Observatory, which is operated by the Association of Universities for Research in Astronomy (AURA) under cooperative agreement with the National Science Foundation.} packages and custom written tools,
following the standard procedures for image reduction. Light curves were
obtained with a median sampling of 2 days in the $B$-band (Johnson band
pass = $4330 \pm 500$\,\AA), $V$-band (Johnson band pass = $5500 \pm
500$\,\AA) and the redshifted H$\beta$ (NB = $5007 \pm 30$\,\AA)
line.
Photometry was performed using a 7$\farcs$5 aperture.
The light curves are calculated relative to $\sim$ 20 nearby non-variables
reference stars located on the same field,
having similar brightness as the AGN.
The absolute calibration was obtained using the measured fluxes of
reference stars from SA095 field
(\citealt{2009AJ....137.4186L})
observed on the same nights as the AGN.

\begin{figure}
  \centering
  \includegraphics[width=\columnwidth]{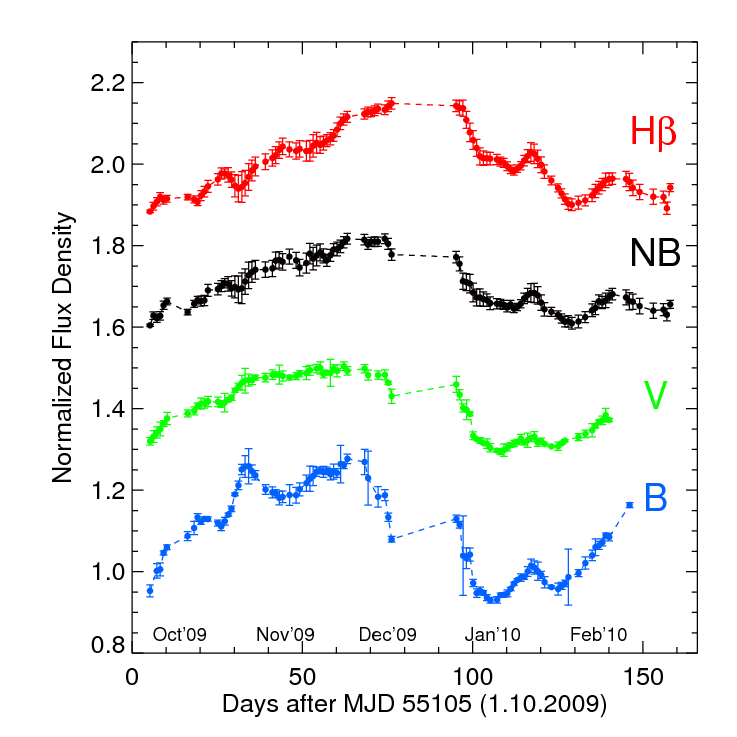}
  \caption{ 
    Observed light curves for \object{3C120} between Oct. 2009 and March
    2010. The H$\beta$ light curve is computed by subtracting a scaled
    $V$ curve from the NB curve and re-normalizing it to mean $ = 1$. 
    The light curves are vertically shifted by multiples of $0.2$ 
    for clarity. The
    data gap between the end of December 2009 and the beginning of January
    2010 is mostly due to strong wind preventing observations.}
\label{lc_3C120}
\end{figure} 

Additionally, we obtained an single epoch spectrum using
the Calar Alto Faint Object Spectrograph (CAFOS) instrument at the 2.2\,m telescope on Calar Alto observatory,
Spain. The spectrograph's slit width was 1$\farcs$54. The reduced
spectrum is shown in Figure~\ref{spectrum_nbfilter}. \object{3C120} lies at redshift $z =0.0331$ so that the H$\beta$ line falls into the NB $5007 \pm 30\AA$ filter. The NB filter effectively covers
the line between velocities -3800km/s and +2200km/s. We determined that about 95\% of the line flux is contained in the NB filter (red shaded line in Figure~\ref{spectrum_nbfilter}) through line profile deconvolution. 

The characteristics 
of the source and a summary of the photometry results are listed in Table~\ref{table1} and Table~\ref{table2} respectively. Note that the magnitudes and the mean of the total flux have been corrected for galactic foreground extinction according to \citet{1998ApJ...500..525S}. 
Observed fluxes in all bands are listed in Table 5.

\section{Results and discussion}

\subsection{Light curves and BLR size}

\begin{table*}
\begin{center}
\caption{Photometry results of 3C120}
\label{table2}
\begin{tabular}{@{}cccccccc}
\hline\hline
$B$ & $V$ & $OIII$ & $fB_{total}^{(1)}$ & $fV_{total}^{(1)}$& $fOIII_{total}^{(1)}$ \\
(mag) & (mag) & (mag) & (mJy) & (mJy) & (mJy) \\
\hline
14.27-14.66 & 13.92-14.18 & 13.09-13.42 & 7.02$\pm$0.11 & 9.27$\pm$0.10 & 18.03$\pm$0.26 \\
\hline
\end{tabular}
\end{center}
\tablefoottext{1}{$fB_{total}$ , $fV_{total}$ and $fOIII_{total}$ refer to the mean of the total flux ranges during our monitoring.}
\end{table*}

The light curves of \object{3C120} are shown in Figure~\ref{lc_3C120}. The
$B$-band shows a
gradual increase from the beginning of October to the beginning 
of December by 30\%.
Afterwards, the flux undergoes an abrupt drop by about 20\% until
mid-December 2009. Between the end of January and early February, the
variability is more regular and the flux increases again to reach a
third maximum at the end of February 2010. In contrast to the steep $B$
band flux increase, the narrow band (NB) 
flux increase is stretched until December
2009 to its first maximum. The sharp $B$ band flux decrease  
in December 2009, is reflected in the NB in
early January 2010. 
Thus, at a first glance, the time delay of the H$\beta$ 
line against the continuum variations is 20 -- 30 days.

\begin{figure}
  \centering
  \includegraphics[width=\columnwidth]{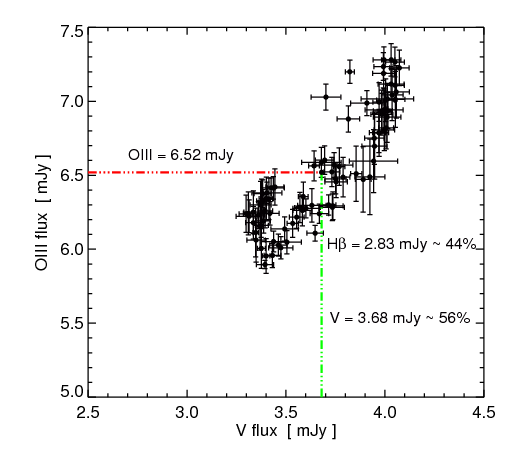}
  \caption{
    Flux-flux diagram for the NB (OIII) and $V$ filter.
    Black dots denote the measurement pair of each night.
    The red and green lines represent the average flux in 
    the OIII and $V$ band respectively. 
    Fluxes were measured using circular 7.5" apertures.
    The data are as observed and not corrected for extinction. 
  }
\label{frac_3c120}
\end{figure}

\begin{figure}
  \centering
  \includegraphics[width=\columnwidth]{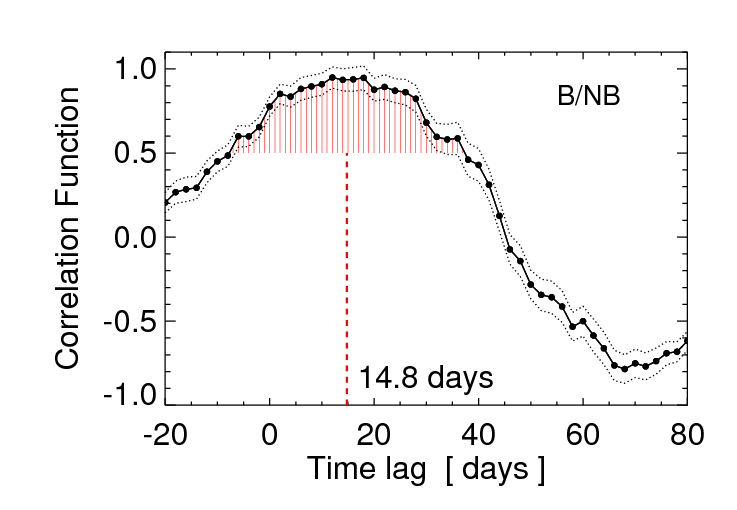}
  \includegraphics[width=\columnwidth]{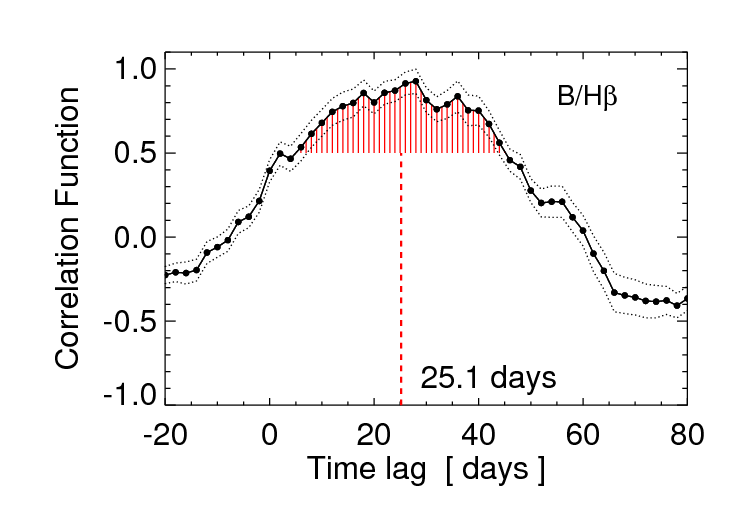}
  \caption{ 
    Cross  correlation of $B$ and NB
    light  curves (top) and  of $B$ and H$\beta$
    light  curves (bottom). 
    The  dotted lines  indicate  the error  range ($  \pm
    1\sigma$) around the cross correlation. 
    The red shaded area marks the range
    used to calculate the centroid of the lag 
    (vertical red dashed line).
  }
\label{DCF_3c120}
\end{figure}

As usual, the precise lag determination is done via cross 
correlation of light curves.
However, as pointed out by Haas et al. (2011), the NB 
contains not only the emission line flux, but also a 
contribution of the varying continuum. 
Then cross correlating $B$ and NB, in principle,  
will result in two peaks, 
one peak from the emission line lag 
and 
one peak at zero lag from the auto-correlation of the continuum. 
These two peaks can only be discerned, 
if their separation is larger than the width of the auto-correlation. 
As illustrated in the top panel of Figure~\ref{DCF_3c120}, in our case 
of \object{3C120} the two peaks are not separated, rather the 
$B/NB$ cross correlation shows only a very broad distribution. 
As proposed by \citet{2012ApJ...747...62C}, 
one could try to disentangle the peaks in the correlation domain, 
for instance by subtracting a $B/V$ autocorrelation from the 
$B/NB$ cross correlation.
But this requires a very precise normalization of the correlations. 
Therefore, we prefer a more straight forward approach by directly removing 
the continuum contribution from the NB light curve 
before applying the correlation techniques.

From the spectrum we estimate that about 50\% continuum and
50\% H$\beta$ emission line contributes to the narrow band
filter.
The spectrum, however, was obtained with a much smaller aperture 
(slit width 1$\farcs$54) than the photometric light curves 
(7$\farcs$5 diameter). Moreover, we have only one single epoch spectrum, 
which does not allow us to easily determine the varying continuum 
contribution to the NB light curve. 
Therefore we consider the NB and V band light curves.
Figure~\ref{frac_3c120} shows for each night 
the narrow band (OIII) and $V$ band fluxes, calibrated to mJy.  
If the flux errors of each night are sufficiently small, 
then in principle one could subtract the $V$ band mJy light curve 
from the NB mJy light curve to obtain the H$\beta$ mJy light curve.
However, our experience shows that this results in a rather 
noisy H$\beta$ light curve. 
Therefore, we used an average scaling of the NB and V band light curves, 
which turned out to yield good results. 
In fact, the results, i.e. the derived lags, 
do not depend sensitively on the precise choice of the 
average scaling factor.

The $V$ band flux, on average, corresponds to $\sim$ 56\% of the 
narrow band flux (Fig.~\ref{frac_3c120}). Considering that the
$V$ flux comprises the contributions from the continuum, 
the H$\beta$ and the OIII
lines, our choice of 50\% continuum flux in the NB is
justifiable.
Thus, we computed a
synthetic H$\beta$ light curve by subtracting a scaled $V$ curve from
the NB curve, i.e. H$\beta$ = NB $-$ 0.5 $V$. 

We used the discrete
correlation function (DCF, \citealt{1988ApJ...333..646E}) 
to cross correlate the light curves. 
The cross correlation of $B$-band and H$\beta$ yields a 
time delay of 25.1 days
defined by the centroid $\tau_{cent}$ 
(Fig.~\ref{DCF_3c120}).\footnote{ 
While a correlation level of $r \geq 0.8r_{max}$ 
is widely used in spectroscopic 
reverberation measurements, well sampled data allow 
to use also lower values of $r$ 
(see Appendix in \citealt{2004ApJ...613..682P}). 
We here calculated the DCF centroid above the 
correlation level at $r \geq 0.5r_{max}$, 
finding that this yields more robust results 
for our data of \object{3C120}, 
presumably because the DCF is quite broad. 
}

\begin{figure}
  \centering
  \includegraphics[width=\columnwidth]{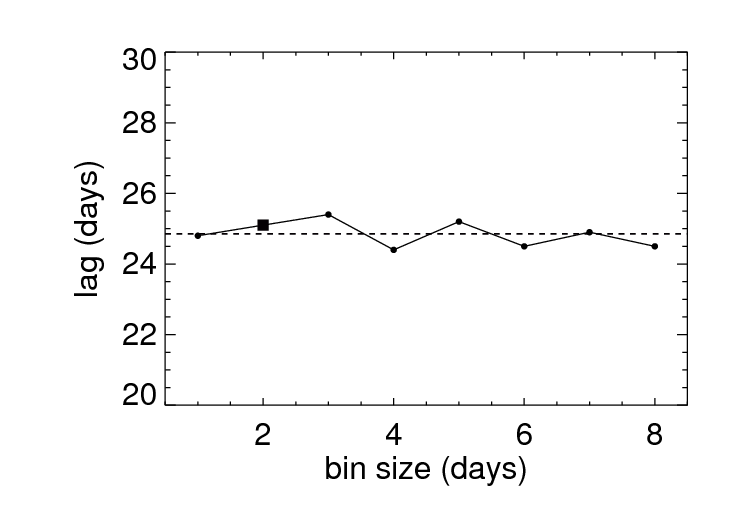}
  \caption{Lag determined by the DCF versus bin size. 
    The square symbol indicates the value used in this work 
    (bin size of 2 days) and 
     the dashed line represents the average value of 24.85.
     The data points are here connected by a line 
     to guide the eye.}
\label{DCF_bin_3c120}
\end{figure}

As usual we adopted the median sampling 
value (2 days) for the bin size in the DCF. 
In addition we checked whether the lag depends on the bin size.
As discussed in detail by Rodriguez-Pascual et al. (1989), 
such dependencies could be caused by the scatter of the points 
in the light curves (bin size smaller than the median sampling)
or by systematic noise structures 
(bin size grater than the median sampling). 
Figure~\ref{DCF_bin_3c120} shows the lag as a function of bin size. 
Any deviation is clearly less than 2\%, arguing in favour of 
our adopted bin size of 2 days.

\begin{figure}
  \centering
  \includegraphics[height=6.5cm,width=\columnwidth]{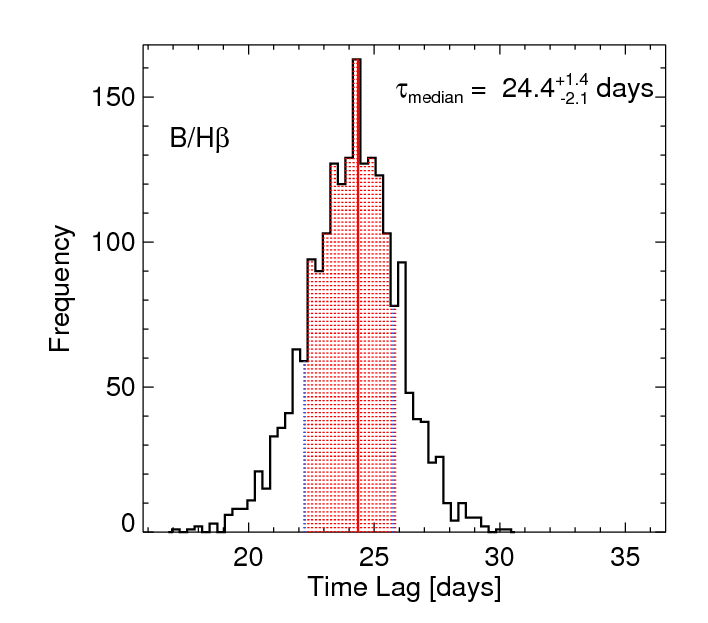}
  \caption{ Results of the lag error analysis. The histogram shows the
    distribution of the centroid lag obtained by cross correlating 2000
    flux randomized and randomly selected subset light curves 
    (FR/RSS method). The red
    area marks the 68\% confidence range used to calculate the errors of
    the centroid.}
  \label{FR_RSS_3c120}
\end{figure}

To determine the uncertainty in the time delay
we applied the flux randomization and random 
subset selection method
(FR/RSS, \citealt{1998PASP..110..660P}, 
\citealt{2004ApJ...613..682P}). 
From the observed light curves we 
create 2000 randomly selected subset
light curves, each containing 63\% of 
the original data.\footnote{Considering the Poisson 
probability of not selecting any particular 
point, reduces
the size of the selected sample by a factor 
about 1/e, resulting in 63\% of the original data 
(\citealt{2004ApJ...613..682P}).
} 
The flux
value of each data point was randomly altered 
consistent with its
(normal-distributed) measurement error. 
We calculated the discrete correlation function for
the 2000 pairs of subset
light curves and the corresponding centroid 
(Fig.~\ref{FR_RSS_3c120}).
This yields a median lag $\tau_{cent}$ = 24.4 $^{+1.4}_{-2.1}$
days. Note the small lag uncertainty of less than 10\%. 
Correcting for the time dilation factor we
obtain a rest frame lag $\tau_{rest}$ = 23.6 $\pm 1.7$ days.
This lag is somewhat smaller 
than the lag $\tau_{rest}$ = 
27 $\pm 1$ days obtained by \citet{2012arXiv1206.6523G}, 
from their one year later spectroscopic 
reverberation mapping campaign. 
We will see below, that the lag difference nicely fits to 
the luminosity difference of \object{3C120} 
at the two campaigns (Sect.~\ref{section_r_l_relation}).

\subsubsection{Virial mass of the central black hole}
\label{section_virial_mass}

Using the velocity dispersion of the emitting gas together with
the BLR size, 
it is possible to determine the mass of the central black hole
following the virial theorem: 
\begin{eqnarray}
{M_{BH}} = f \frac{R \cdot \sigma_{V}^2}{G}
\end{eqnarray}
where $\sigma_{V}$ is the emission-line velocity dispersion (assuming
Keplerian 
orbits of the BLR clouds), $R=c \cdot \tau$ is the BLR size and the
factor $f$ depends on the geometry and kinematics of the BLR
(\citealt{2004ApJ...613..682P} and 
references therein). Most of the results presented in previous
reverberation studies have been carried out  
considering only the virial product $c \tau \sigma_{V}^2/G$,
ie. assuming a scaling factor $f=1$.

The first empirical calibration of the factor $f$ was obtained by
\citet{2004ApJ...615..645O} who determined an average value of
5.5$\pm$1.8,  
assuming that AGNs and quiescent galaxies follow the same
$M_{BH}-\sigma_{*}$ relationship. On the other hand, studies by \citet{2006A&A...456...75C} show that the H$\beta$ profile is narrower in the rms spectra than in the mean spectra. Therefore, a higher velocity dispersion is expected in the mean spectra. Hence, they suggest a scaling factor $f = 3.85$ (his table 2). Subsequents studies by
\citet{2011MNRAS.412.2211G} 
(based on the method established by \citealt{2004ApJ...615..645O})
suggest a different scenario, reducing this first value to half
(2.8$\pm$0.6).   
Therefore, the determination of the factor $f$ is an open issue
involving theoretical models and empirical
relationsh.png. Nevertheless,  
we here calculated the black hole mass ($M_{BH}$) adopting the value
obtained by \citet{2004ApJ...615..645O}, in order to compare our
results with the literature. 

\begin{figure}
  \centering
  \includegraphics[width=\columnwidth]{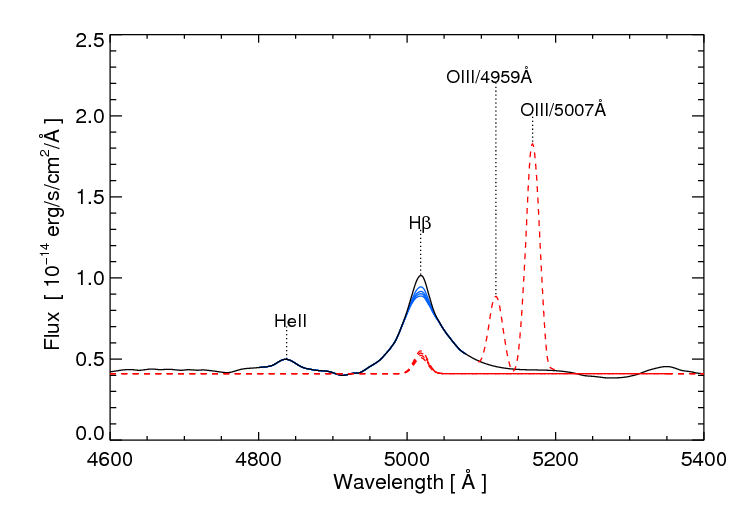}
  \caption{
      CAFOS spectrum of \object{3C120}, zoomed onto the H$\beta$ line.
      The solid black curve represents the spectrum after
      subtracting the narrow [OIII]$\lambda\lambda$4959,5007 emission
      lines (red dashed lines). 
      The four dashed red lines (at the bottom of the H$\beta$ profile) show the narrow H$\beta$ models, for four
      intensities between 7\% and 10\% of the [OIII]$\lambda$5007 line.
      The four solid blue lines 
      represent the spectrum after subtracting the narrow H$\beta$ models.
    Note that the broad faint 
    HeII$\lambda$4686 emission line can be clearly seen, but its
    possible contribution to the H$\beta$ profile is negligible.
  } 
\label{narrow_subtracted}
\end{figure}

The velocity dispersion is determined from the width of the
emission-line. It can be characterized by the second moment of the
line profile ($\sigma_{\rm line}$) or by its full width at half
maximum (FWHM). 
The first appears to be the most robust against measurement
uncertainties (\citealt{2004ApJ...613..682P}) and is widely used in
subsequent reverberation studies. 
Usually - for spectroscopic monitoring data - the line dispersion
$\sigma_{\rm line}$ is measured from the root-mean-square (rms)
spectrum, in which the contribution of the narrow emission lines
components disappears. 

The feasibility of the use of single-epoch (SE) spectra for the black
hole mass determination has been established in previous 
investigations (e.g. \citealt{2002ApJ...571..733V},
\citealt{2007ApJ...661...60W}, \citealt{2009ApJ...692..246D}). 
On average, uncertainties of $\sim$30\% have been 
reported for black hole mass determination from single epoch spectra
measurements.
However, the results of these studies have detected systematic sources
of uncertainties, which arise mainly from the emission-line velocity
dispersion and narrow line contamination (see
\citealt{2009ApJ...692..246D}, for details). 

We here try to determine the black hole mass of 3C120 from our BLR
size and the contemporaneous single epoch CAFOS spectrum.

The narrow emission lines [OIII]$\lambda\lambda$4959,5007 were modeled
by single Gaussian profile and subtracted from the (SE)
spectra (following \citealt{2009ApJ...692..246D}). We also considered the
theoretical intensity ratio for this doublet of 2.98 determined by
\citet{2000MNRAS.312..813S} which is consistent with observational
measurements and exhibit a negligible intensity variation
(\citealt{2007MNRAS.374.1181D}). 
A challenge is to
  remove the narrow component of the H$\beta$ profile. 
  Under the assumption of standard emission-line ratios,
  we model this contribution
  (following \citealt{2009ApJ...692..246D}). 
  The [OIII]$\lambda$5007 line profile was
  used to create the model of the narrow H$\beta$ emission, 
  adopting a narrow H$\beta$ strength between 7\% and 10\% of the
  [OIII]$\lambda$5007 strength
  (following \citealt{2007ApJ...661...60W}). 
  This narrow H$\beta$  model was then scaled and
  subtracted from the observed H$\beta$ line profile.
Figure~\ref{narrow_subtracted} shows the original H$\beta\lambda$4861
profile together with the narrow emission lines subtracted. 

After removing the narrow H$\beta$ line (and the [OIII] lines), the
H$\beta$ profile was isolated by the subtraction of a linear continuum fit, obtained
through interpolation between two continuum segments on either end of the line,
taking into account the possibility of local continuum contamination by FeII contribution
and the red wing of the HeII$\lambda$4686 emission line (which may be blended with the blue
wing of H$\beta$). Fig.~\ref{narrow_subtracted} shows the original H$\beta\lambda$4861 profile together
with the narrow emission lines subtracted. 
The velocity dispersion after removal of narrow lines is $\sigma_{\rm line} = 1504\, km/s$.

From their recent spectroscopic
  reverberation campaign, \citet{2012arXiv1206.6523G} determined
  $\sigma_{\rm line}(mean) = 1687 \pm 4\, km/s$ and $\sigma_{\rm
    line}(rms) = 1514 \pm 65\, km/s$. 
  Note that in their mean spectra  (their Fig. 1) the narrow
  H$\beta$ component appears stronger than in our spectrum, 
  suggesting that our
  velocity dispersion may be underestimated. If true, this would be
  one of the
  possible systematic effects of single epoch spectra
  (\citealt{2009ApJ...692..246D}). On the other hand, our time lag is
  about 40\% smaller than the result derived by
  \citet{2004ApJ...613..682P}. Hence, according to photo-ionization
  physics, during our
  campaign the H$\beta$ region is closer to the ionizing source and in
  consequence higher velocities of the gas clouds are expected.

One shoud keep in mind that the  black hole mass determination relies on the
assumption that the BLR emitting gas clouds are in
virialized motion. Although this has been shown for a large
sample of AGNs, there are a few exceptions that have emerged as a
fundamental limitation for black hole mass determination via 
reverberation mapping
(e.g. \citealt{2009ApJ...704L..80D},
\citealt{2010ASPC..427...68G}). Single epoch spectra can not identify
such different velocity signatures of
H$\beta$ clouds with a complex kinematical behavior. 

Using the previous value (with 25\% uncertainty adopted and our rest frame time
lag $\tau = 23.6$d, 
the virial black hole mass is $M_{virial} = (10 \pm 5) \times
10^{6} M_{\odot}$ which is consistent, inside the margins of errors, with 
$M_{virial} = (12 \pm 1) \times 10^{6} M_{\odot}$ derived by \citet{2012arXiv1206.6523G} 
from their recent spectroscopic monitoring campaign and with $M_{virial} = (10 \pm 5) \times 10^{6} M_{\odot}$
derived via spectroscopic reverberation mapping by \citet{2004ApJ...613..682P}.
Considering the factor $f=5.5 \pm 1.8$, we determine a central black hole mass $M_{BH} =(57 \pm 27) \times 10^{6} M_{\odot}$.

\subsubsection{Host-subtracted AGN luminosity}
\label{section_agn_luminosity}

In order to determine the pure AGN luminosity, commonly at 5100\AA,
the contribution of the host galaxy to the nuclear flux has to be 
subtracted.
The contribution of the host galaxy to the nuclear flux of 3C120 
has been studied by Bentz et al. (2006, 2009a) using high 
resolution HST imaging as well as by \citet{1992MNRAS.257..659W}, 
\citet{1997MNRAS.292..273W} and \citet{2010ApJ...711..461S} using the so-called flux 
variation gradient method, which was originally proposed by 
\citet{1981AcA....31..293C}.
We here apply the FVG method to our \object{3C120} data and compare our results 
with the previous ones. Because the FVG method appears to be not 
widely known, we start with some comprehensive explanations here.

~\\
{\it \noindent 3.1.2.1 On the flux variation gradient method}
~\\

For multi-band AGN monitoring data, 
Choloniewski noticed a linear relationship between the 
fluxes obtained in two differents filters taken at 
different epochs. Using this relationship he demonstrated the 
existence of two component structure to the spectra of Seyfert galaxies. 
The two component structure assumes that one component (the host) 
is constant in time and that the second component (the AGN) 
has a strong variability. 
Despite of this strong variability the spectral energy 
distribution (SED) of the AGN does not change. 
Choloniewski plotted the observed UVB fluxes for 40 Seyfert 
galaxies as 3 dimensional vectors. 
This gave a clear geometrical interpretation of the total flux 
variations and allowed for the decomposition of the total vector flux $\vec \phi(t)$ into 
a constant vector $\vec F$ and a variable vector $\vec f(t)$ (his figure 1). 

Following the geometrical representation from \citet{1981AcA....31..293C}, is possible to write the two component structure using two different filters $(i,j)$ as:
\begin{eqnarray}
{\vec \phi_{i,j}(\nu,t)} = {\vec F_{i,j}(\nu)}+{\vec f_{i,j}(\nu,t)}
\end{eqnarray}
where the observed flux $\vec F_{i,j}(\nu)$ and $\vec f_{i,j}(\nu,t)$ are the constant and variable components in two arbitrary filters respectively.

With the - at optical wavelengths 
observationally corroborated - assumption that the 
shape of spectrum of the variable component 
does not change, one deduces:
\begin{eqnarray}
{{\vec f_{j}(\nu,t) \over \vec f_{i}(\nu,t)}} = a_{ji}
\end{eqnarray}
where $a_{ji}$ is a constant. Using equations (2) and (3) one obtains 
the equation of a straight line in the plane 
($\vec \phi_{i}$, $\vec \phi_{j}$):
\begin{eqnarray}
{\vec \phi_{j}(\nu,t)} = {a_{ji}} {\vec \phi_{i}(\nu,t)}+{b_{ji}}
\end{eqnarray}
where ${b_{ji}}$ = ${\vec F_{j}(\nu)} - {a_{ji}} {\vec F_{i}(\nu)}$

Thus the coefficient $a_{ji}$ represents the color index 
of the variable component. This coefficient of proportionality 
has also been called the flux variation gradient (FVG). 
It is denoted by the symbol $\Gamma$ by \citet{1992MNRAS.257..659W}. 
Both, $a_{ji}$ and $b_{ji}$ coefficients have to be determined 
by a linear regression analysis.

The choice of which regression method to use 
depends on the degree of knowledge about the data, 
especially on how well separated the dependent and 
independent variables are, the measurement errors, 
the intrinsic dispersion of data around the best 
fit and others factors (see \citealt{1990ApJ...364..104I}, for details).

\begin{figure}
  \centering
  \includegraphics[width=\columnwidth,clip=true]{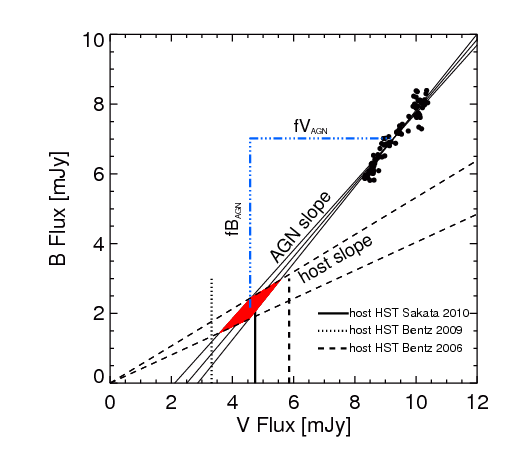}
  \caption{ Flux variation gradient diagram of \object{3C120}. 
    The solid lines delineate the
    bisector regression model yielding the range of the AGN slope. 
    The
    dashed lines indicate the range of host slopes determined
    by \protect\citet{2010ApJ...711..461S} for 11 nearby AGN. 
    The intersection between the host galaxy and AGN slope (red area)
    gives the host galaxy flux in both bands. The two vertical
    dotted and dashed
    lines show the host flux determined by \protect\citet{2006ApJ...644..133B}
    and \protect\citet{2009ApJ...697..160B} respectively. 
    The solid line shows the host flux obtained by \protect\citet{2010ApJ...711..461S}. 
    The dash-dotted
    blue lines represent the range of the AGN flux in both
    filters. Fluxes (black dots) were measured in a 7.5" aperture
    photometry and corrected for galactic foreground extinction.  }
  \label{3C120_FVG}
\end{figure}

\citet{1992MNRAS.257..659W} 
proposed a new method to separate the nuclear flux
from the host galaxy, based on the flux-flux diagrams by
\citet{1981AcA....31..293C}. 
Winkler et al. monitored 35 southern Seyfert galaxies using 
UBVRI multi-aperture photometry to estimate the colors 
of the host galaxy. They plotted the total fluxes obtained 
through 20$^{\prime\prime}$ and 30$^{\prime\prime}$ apertures, 
together with a line representing the fluxes from a 
component having the same colors as the host galaxy. 
The intersection of this line with the linear regression 
fit obtained from the total flux represents then the 
contribution from the host galaxy. 
As the observed source varies in
luminosity, the fluxes in the FVG diagram will follow a 
slope representing the AGN color while the host will 
show no variation. The nuclear flux is then calculated 
by subtracting the constant host galaxy component 
(obtained by the FVGs method) from the total flux.

The studies by \citet{1981AcA....31..293C} and 
\citet{1992MNRAS.257..659W} observationally established 
at optical wavelengths like in the BVRI bands 
that the AGN and host colors, i.e. flux ratios, are
different but stay constant with time.
Although these results have been corroborated by numerous authors
(e.g., \citealt{2003AAS...202.3803Y}; \citealt{2006ApJ...639...46S}; \citealt{2008ARep...52..167D}; \citealt{2010ApJ...711..461S}), 
they are still a matter of a debate. 
For a more extensive discussion we refer the reader 
to \citealt{2010ApJ...711..461S}.

A valuable study on the determination of the nuclear flux contribution was conducted by \citet{2010ApJ...711..461S}, using both HST imaging and the FVG method. In this study the flux of the host galaxy was estimated for 11 nearby Seyfert galaxies and QSOs. These fluxes were measured in \textit{B}, \textit{V} and \textit{I} bands by using surface brightness fitting to the high resolution Hubble Space Telescope (HST) images and from the MAGNUM observations. The authors find a well defined range for the host galaxy slope of $0.4 < \Gamma^{host}_{BV} < 0.53$, which corresponds to host galaxy colors of $0.8 < B-V < 1.1$. The results - in particular on \object{3C120} - obtained by \citet{2010ApJ...711..461S} are consistent (within measurement errors) with those obtained by \citet{1997MNRAS.292..273W}, \citet{2006ApJ...644..133B} and \citet{2009ApJ...697..160B}.

~\\
{\it \noindent 3.1.2.2 Application to \object{3C120}}
~\\

Figure \ref{3C120_FVG} shows the $B$ and $V$ fluxes 
obtained during the same nights
and through the same aperture in a flux-flux diagram. 
The host color range is taken from \citet{2010ApJ...711..461S} 
and drawn from the origin of ordinates (dashed lines). For the varying
AGN flux (black dots), we use linear regression in order to determine a
range of possible AGN colors (solid thin lines). The cross-section of
the host and AGN slope allows us now to split the superposition of
fluxes in both filters. 
Note that the $B$ and $V$  fluxes have been corrected for
galactic foreground extinction according to \citet{1998ApJ...500..525S}.

In both filters, 
the total flux (AGN+Host) contains a contribution from the emission lines originating
in the narrow line region (NLR). However, this contribution is less than
10\% of the host galaxy flux in the $B$ and $V$ band
(\citealt{2010ApJ...711..461S}). 
We here define the host galaxy to include the NLR line contribution.

Flux Variations Gradients (FVGs) were evaluated by fitting a straight
line to the data using five methods of linear regression; OLS(Y/X),
OLS(X/Y), OLS bisector, Orthogonal regression and Reduced Major-Axis (RMA)
were used, depending on the corresponding error treatment of each
method.
Our regression algorithms are based on the formulas 
of \citet{1990ApJ...364..104I} (their Table 1) and the
variance for each method was calculated using the same formulation. 
The statistics of each linear regression fit are listed in 
Table~\ref{table3}.

\begin{table}
\caption{Linear regression results.}
\label{table3}
\begin{tabular}{lcccccc}
\hline
Method & $a^{a}$ & $\sigma_{a}$ & $b^{b}$ & $\sigma_{b}$ &&\\
\hline
OLS(Y/X) & 1.04  &0.04  & -2.60 & 0.53 &&\\
OLS(X/Y) & 1.21  &0.05  & -4.22 & 0.33 &&\\
OLS Bisector & 1.12  &0.04  & -3.38 & 0.39 &&\\
Orthogonal Regression & 1.13  &0.05  & -3.48 & 0.41 &&\\
RMA & 1.22  &0.06  & -4.18 & 0.49 &&\\
\hline
\end{tabular}
\tablefoottext{a}{Slope.}\\
\tablefoottext{b}{Intercept coefficient.}\\  
\end{table}

The bisector linear
regression line and the OLS(X/Y) method yields a linear gradient of $\Gamma_{BV} = 1.12 \pm
0.04$ and $\Gamma_{BV} = 1.04 \pm
0.04$ respectively. The results are consistent, within the uncertainties,
with $\Gamma_{BV} = 1.11 \pm 0.02$ determined by \citet{2010ApJ...711..461S} and 
$\Gamma_{BV} = 1.02 \pm 0.07$ obtained
by \citet{1997MNRAS.292..273W}. 
While the uncertainties to the OLS bisector and OLS(Y/X) regression slopes 
are lower (about 4\%). According to \citet{1990ApJ...364..104I}, the OLS bisector method is the most
suitable in order to determine the underlying relationship between the variables and it was also considered in previous studies of FVG method
by \citet{1997MNRAS.292..273W}. 
Therefore, we adopted the OLS bisector linear regression to define the range of AGN slope.
 
Averaging over the intersection area between the AGN and the host galaxy
slopes, we obtain a mean host galaxy flux of $(2.17 \pm 0.33)$ mJy
in $B$ and $(4.58 \pm 0.40)$ mJy in $V$. 
Our host galaxy flux derived with the FVG method is 
consistent with the values $f_B \approx{2.10} $ mJy
and $f_V \approx{4.73}$ mJy obtained by \citet{2010ApJ...711..461S}. 
Note that one has to coadd the values of Tables 5 and 8 of Sakata et al., 
in order to include the flux contribution of the narrow lines 
([OIII]$\lambda$4959, 5007, H$\beta$, H$\gamma$) to each filter. 

We have also compared our results with those 
obtained by \citet{2006ApJ...644..133B}
and \citet{2009ApJ...697..160B} through modeling of the 
host galaxy profile (GALFIT,
\citealt{2002AJ....124..266P}) from the high-resolution 
HST images. 
On the nucleus-free image and through
an aperture of $5\arcsec \times 7\farcs6$, 
\citet{2006ApJ...644..133B} determined a rest-frame 5100\AA~ host-flux
$F_{5100\AA~}$ =$1.82\times10^{-15} erg s^{-1} cm^{-2} \AA^{-1}$. 
Using the color term factor from the HST-F550M filter to 5100\AA~(their Table 3)
we deduced the flux for the F550M filter to be 
$F_{F550M}$ =$1.74\times10^{-15} erg s^{-1} cm^{-2} \AA^{-1}$. 
This corresponds after exinction 
correction to $4.62 mJy$. 
Moreover, considering the contribution of 
the narrow lines in the $V$-band ($1.235 mJy$), 
determined by \citet{2010ApJ...711..461S}, the
previous value translates to $5.86 mJy$.
In a subsequent investigation, \citet{2009ApJ...697..160B} 
determined a host galaxy flux of
$0.78\times10^{-15} erg s^{-1} cm^{-2} \AA^{-1}$, 
which after extinction correction and adding the contribution 
of the narrow lines yield a value of $3.31 mJy$.
The difference between the two results by Bentz et al. lies mainly 
in the type of model used for the decomposition of the galaxy. 
The first study 
considered a general Sersic function for modeling bulges 
(\citealt{2006ApJ...644..133B}). 
The second study performed the modeling with variations 
and improvements to the original profile 
(\citealt{2009ApJ...697..160B}). 
Our value ($fV_{host} = 4.58 mJy$) falls exactly in the middle between 
both values, simply suggesting that our determination is consistent, 
within the error margins, with those determined by 
Bentz et al.\footnote{
The actual debate on the  host flux discrepancies cannot be solved here. 
We just note that 3C120 (together with another three objects) shows 
larger residuals in the surface brightness fit of the MAGNUM $V$-band images 
(Fig.~17 of \citealt{2010ApJ...711..461S}) compared to the HST images 
(Fig.~14 of \citealt{2010ApJ...711..461S} and Fig.~3 of \citealt{2009ApJ...697..160B}). 
This may suggest a 
possible host overestimation. 
On the other hand, \citet{2009ApJ...697..160B} mentioned a possible host  
underestimation caused by the uncertain sky background.
Futhermore, 
our aperture area ($7\farcs5$ in diameter) is only 16\% 
larger than that of Peterson's campaign ($5\arcsec \times 7\farcs6$). 
This together with an additional bandpass conversion factor between 
the HST-F550M filter and our Johnson-V filter would 
increase the \citet{2009ApJ...697..160B} host flux by about 10\%.
This suggests that the most contemporaneus value quoted 
by \citet{2009ApJ...697..160B} could be underestimated.
}

\begin{figure}
  \centering
  \includegraphics[width=\columnwidth,clip=true]{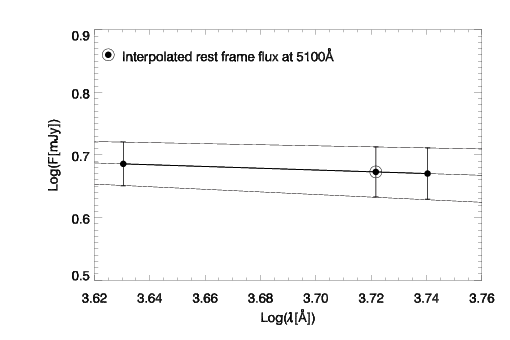}
  \caption{Interpolated rest frame flux at 5100\AA~ from $fB_{AGN}$ and $fV_{AGN}$ fluxes (black dots) for an AGN continuum with power law spectral shape ($F_{\nu} \propto
\nu^{\alpha}$) (solid lines).}
  \label{interpolation}
\end{figure}

The AGN fluxes at the time of our monitoring can be
determined by subtracting the host galaxy fluxes from the total
fluxes. During our monitoring campaign, the total $B$ fluxes lie in the
range between 5.81-8.39 mJy with a mean of $(7.02 \pm 0.11)$ mJy. The
total $V$ fluxes lie in the range between 8.17-10.38 mJy with a mean
$(9.27 \pm 0.11)$ mJy. The host galaxy subtracted average AGN fluxes and
the host galaxy flux contribution of \object{3C120} are listed in
Table~\ref{table4}.

Also listed in Table~\ref{table4} are the interpolated rest frame
$5100\AA~$ fluxes and the monochromatic AGN luminosity $\lambda
L_{\rm{\lambda(AGN)}}$ at $5100\AA~$. 
The rest frame flux at 5100\AA~
was interpolated from the host-subtracted AGN fluxes in both bands, adopting for the 
interpolation that the AGN has 
a power law SED ($F_{\nu} \propto
\nu^{\alpha}$) with an spectral 
index $\alpha=\log(fB_{AGN} / fV_{AGN}) / \log(\nu_{B} / \nu_{V})$,
where $\nu_{B}$ and $\nu_{V}$ 
are the effective frequencies in the B and V bands, respectively. 
The error was determined by 
interpolation between the ranges of the AGN fluxes $\pm\sigma$ 
in both filters, respectively, as illustrated in Fig.~\ref{interpolation}.

To determine the luminosity, we used the distance of 145 Mpc
(\citealt{2009ApJ...697..160B}). This yields $L_{\rm{AGN}} = (6.94 \pm 
0.71)\times 10^{43}erg s^{-1}$. The AGN luminosity during our campaign is about
half the mean value of $(12.40 \pm 2.6)\times10^{43}erg
s^{-1}$ found by \citet{2009ApJ...697..160B} 
for their campaign about 15 years ago 
and slightly lower than the value 
$(9 \pm 1)\times10^{43}erg s^{-1}$ found 
by \citet{2012arXiv1206.6523G} for their campaign one year after ours.

\begin{figure*}
  \centering
  \includegraphics[width=18cm,clip=true]{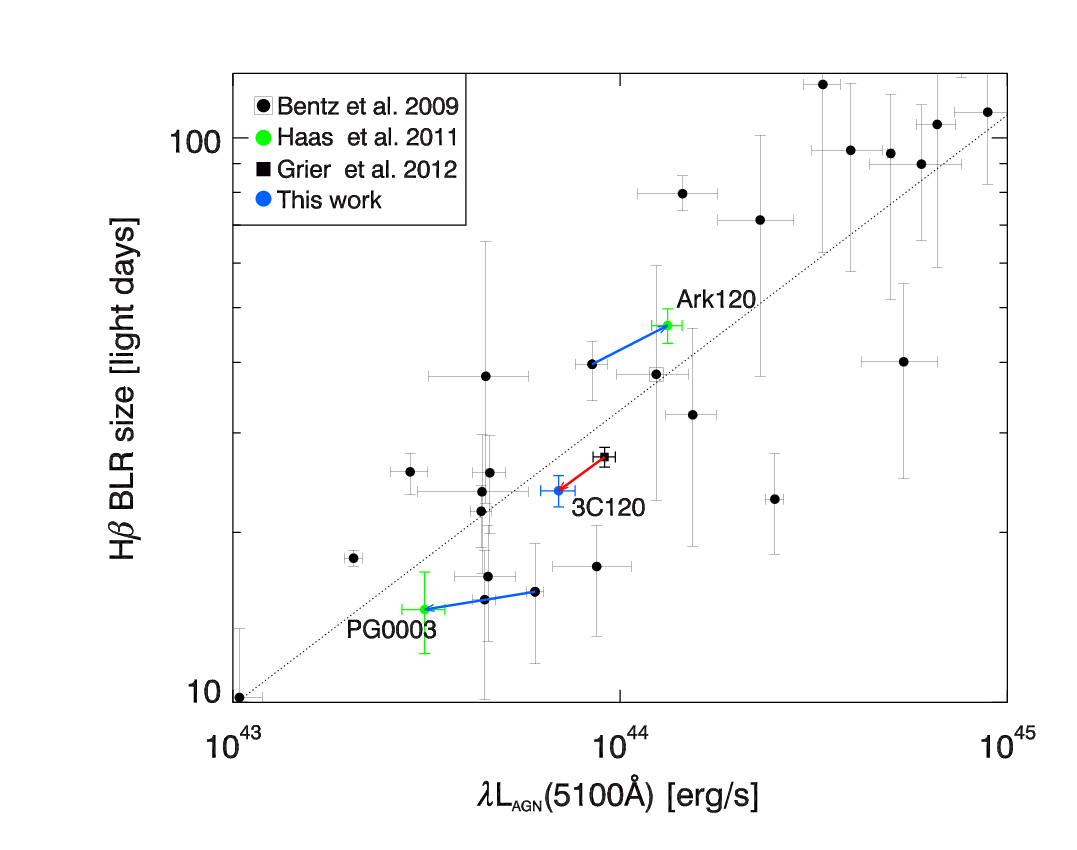}
  \caption{ 
    $R_{BLR} - L$ relationship from data of \protect\citet{2009ApJ...697..160B} (black dots)
    with a fitted slope $\alpha = 0.519$ (dotted line).
    The diagram is zoomed to contain the objects of this work (\object{3C120}, blue dot) and
    \protect\citet{2011A&A...535A..73H} (\object{Ark120} and \object{PG0003}, green dots). 
    The blue arrows show the positional shift of the new measurements 
    with respect to the previous ones from \protect\citet{2009ApJ...697..160B}. 
      The red arrow shows the shift of \object{3C120} between our result and that 
      obtained by \protect\citet{2012arXiv1206.6523G}. 
      The slope $\alpha$ of this shift is remarkably close to the theoretically 
      expected value of $\alpha = 0.5$. 
      The original position of \object{3C120} by Bentz et al. is in the center
      of the plot and marked by the square 
      surrounding the fat dot.
  }
  \label{R_BLR}
\end{figure*}

\subsubsection{The BLR size - luminosity relationship}
\label{section_r_l_relation}

From spectroscopic reverberation mapping, 
the relationship between
the H$\beta$ BLR size and the luminosity (5100\AA) 
of the AGN $R_{{\rm BLR}} \propto L^{\alpha}$
has been established by
\citet{2000ApJ...533..631K}. 
\citet{2009ApJ...697..160B} determined an improved 
slope of $\alpha =$ 0.519$^{0.063}_{-0.066}$ from 
several spectroscopic
reverberation mapping campaigns.

Figure~\ref{R_BLR} shows the $R_{{\rm BLR}}$ and $L_{AGN}$ values obtained by
\citet{2009ApJ...697..160B}, 
the result for 3C120 obtained by \citet{2012arXiv1206.6523G} and the two
objects Ark120 and PG0003 studied with well sampled photometric reverberation mapping 
by \citet{2011A&A...535A..73H}. 
While the
relationship appears to be well defined, 
many objects have large uncertainties yet 
and/or lie quite off the regression line. 
Part of this may simply be due to the fact that AGN are complex 
objects, but part of the dispersion may be due to poor early 
reverberation data. We just note that the new position of 
both \object{Ark120} and \object{PG0003} lies closer to the regression line in Fig.~\ref{R_BLR}.

For \object{3C120} we have now three positions in the R-L diagram:
one from Peterson's spectroscopic monitoring 
campaign in 1998, quite on the regression line 
but having large uncertainties presumably due to sparse sampling;
one from the spectroscopic monitoring campaign in 2010/2011 by \citet{2012arXiv1206.6523G}; 
and one from our photometric monitoring campaign in 2009/2010.
The last two campaigns both have a good time sampling and small
uncertainties.
The striking result from these two campaigns 
is that the slope between these positions of \object{3C120} is $\alpha =
0.504$, hence dueing the brightness changes \object{3C120} 
moves parallel to the theoretically expected slope 
(red arrow in Fig.~\ref{R_BLR}).
With increasing luminosity the BLR size grows 
proportional to the square root of the luminosity, 
and these changes appear to occur rather quickly, 
i.e. within days or weeks.
However, this does not mean that the BLR gas clouds move that rapidly 
in- or outwards.  
\citet{1995ApJ...455L.119B} have presented a model of locally optimally 
emitting clouds.

\begin{table*}
\begin{center}
\caption{Host galaxy subtracted AGN fluxes of 3C120}
\label{table4}
\begin{tabular}{@{}cccccc}
\hline\hline
$fB_{host}$ & $fB_{AGN}^{1}$ & $fV_{host}$ & $fV_{AGN}^{1}$ & $f_{AGN} ((1+z)5100$\AA~ & $\lambda L_{\lambda,AGN}5100$\AA~ \\
    (mJy) & (mJy) & (mJy) & (mJy) & (mJy) & ($10^{43} erg s^{-1}$) \\
\hline
2.17$\pm$0.33 & 4.85$\pm$0.35 & 4.58$\pm$0.40 & 4.68$\pm$0.41 & 4.72$\pm$0.40 & 6.94$\pm$0.71 \\
\hline
\end{tabular}
\end{center}
\tablefoottext{1}{AGN fluxes values $fB_{AGN} = fB_{total}-fB_{host}$ and $fV_{AGN}=fV_{total}-fV_{host}$, with uncertainty range $\sigma_{AGN} = (\sigma_{total}^{2} + \sigma_{host}^{2})^{0.5}$.}\\  
\end{table*}

It successfully explains the emission line characteristics 
of quasar spectra. 
This model provides also a nice explanation for the R-L relation.
In the BLR the gas clouds actually populate a large range of radii, 
but, depending on the density and ionisation parameter,
only at a suited narrow distance range (from the illuminating central
power house) the gas clouds efficiently convert the nuclear 
continuum radiation into line emission. 
In this picture, 
a central continuum brightness variation can quickly change 
the BLR size proportional to $L^{0.5}$.

\section{Summary and conclusions}
\label{section_conclusions}

Using a robotic 15 cm telescope located at an excellent site, 
we have performed a five months monitoring campaign for the Seyfert 1
galaxy \object{3C120}. We determined the broad line region
size, the virial black hole mass and the host-subtracted AGN
luminosity.  The results are:

\begin{enumerate}
\item  The time lag $\tau_{rest}$ = 23.6 $\pm 1.69$ days, obtained from
cross correlation of the H$\beta$ emission line with the optical
continuum light curve,
has changed over one decade, but the physical relation $R_{{\rm BLR}} \propto L_{{\rm AGN,\lambda 5100}}^{1/2}$ still holds.
The small uncertainty in our measurements (7\%) is
presumably due to the well sampled photometric reverberation data.

\item The black hole mass $M_{BH} = 57 \pm 27 \times 10^{6}
M_{\odot}$ is consistent with the value of $55.5 \pm
26.9 \times 10^{6} M_{\odot}$ derived by \citet{2004ApJ...613..682P} from spectroscopic reverberation mapping
data and with the value of $M_{BH} = 67 \pm 6 \times 10^{6}$ determined by \citet{2012arXiv1206.6523G}. 
However, the high uncertainties in the black hole mass, with respect to the most contemporaneus result, are attributed to the adopted 25\% 
uncertainty for the line velocity dispersion and also considering the uncertainty introduced by the scale factor $f$.

\item  Using the flux variation gradient method (FVG) and a
  conservatively limited host galaxy color range, it is possible to find
  the host galaxy subtracted AGN luminosity of \object{3C120} at the time of our monitoring campaign to be
  $L_{\rm{AGN}} = (6.94 \pm 0.71)\times 10^{43}erg s^{-1}$.

\end{enumerate}

The new results obtained for the BLR size and AGN luminosity of 
\object{3C120} fit well into the BLR size-luminosity diagram. 
We conclude that photometric reverberation mapping is an
attractive method with the advantage to efficiently measure the 
BLR and the host-subtracted luminosities for large samples of 
quasars and AGNs. Not only applicable with small telescopes, 
photometric AGN reverberation mapping could become a key tool for the upcoming 
large monitoring campaigns, for instance with the 
Large Synoptic Survey Telescope (LSST). 
This could be an important step forward in order to constrain 
cosmologial parameters from the $R_{{\rm BLR}} \propto L_{{\rm
    AGN,\lambda 5100}}^{1/2}$ 
relationship in order to determine quasar distances and to probe 
the dark energy
(\citealt{2011A&A...535A..73H};\citealt{2011ApJ...740L..49W}) 
as well as to enlarge the current statistics for black hole masses.

\begin{acknowledgements}
 
  This publication is supported as a project of the
  Nordrhein-Westf\"alische Akademie der Wissenschaften und der K\"unste
  in the framework of the academy program by the Federal Republic of
  Germany and the state Nordrhein-Westfalen, as well as by the
  CONICYT GEMINI National programme fund 32090025 for the development
  of Astronomy and related Sciences.

  The observations on Cerro Armazones benefitted
  from the care of the guardians Hector Labra, Gerard Pino, Alberto
  Lavin, and Francisco Arraya.

  This research has made use of the NASA/IPAC
  Extragalactic Database (NED) which is operated by the Jet Propulsion
  Laboratory, California Institute of Technology, under contract with
  the National Aeronautics and Space Administration. This research has made 
  use of the SIMBAD database, operated at CDS, Strasbourg, France. 
  We thank the anonymous referee for his comments and careful review of the manuscript.

\end{acknowledgements}

\bibliographystyle{aa} 
\bibliography{3C120_prm}

\begin{thebibliography}{41}
\expandafter\ifx\csname natexlab\endcsname\relax\def\natexlab#1{#1}\fi

\bibitem[{{Baldwin} {et~al.}(1995){Baldwin}, {Ferland}, {Korista}, \&
  {Verner}}]{1995ApJ...455L.119B}
{Baldwin}, J., {Ferland}, G., {Korista}, K., \& {Verner}, D. 1995, \apjl, 455,
  L119

\bibitem[{{Bentz} {et~al.}(2009{\natexlab{a}}){Bentz}, {Peterson}, {Netzer},
  {Pogge}, \& {Vestergaard}}]{2009ApJ...697..160B}
{Bentz}, M.~C., {Peterson}, B.~M., {Netzer}, H., {Pogge}, R.~W., \&
  {Vestergaard}, M. 2009{\natexlab{a}}, \apj, 697, 160

\bibitem[{{Bentz} {et~al.}(2006){Bentz}, {Peterson}, {Pogge}, {Vestergaard}, \&
  {Onken}}]{2006ApJ...644..133B}
{Bentz}, M.~C., {Peterson}, B.~M., {Pogge}, R.~W., {Vestergaard}, M., \&
  {Onken}, C.~A. 2006, \apj, 644, 133

\bibitem[{{Bentz} {et~al.}(2009{\natexlab{b}}){Bentz}, {Walsh}, {Barth},
  {Baliber}, {Bennert}, {Canalizo}, {Filippenko}, {Ganeshalingam}, {Gates},
  {Greene}, {Hidas}, {Hiner}, {Lee}, {Li}, {Malkan}, {Minezaki}, {Sakata},
  {Serduke}, {Silverman}, {Steele}, {Stern}, {Street}, {Thornton}, {Treu},
  {Wang}, {Woo}, \& {Yoshii}}]{2009ApJ...705..199B}
{Bentz}, M.~C., {Walsh}, J.~L., {Barth}, A.~J., {et~al.} 2009{\natexlab{b}},
  \apj, 705, 199

\bibitem[{{Blandford} \& {McKee}(1982)}]{1982ApJ...255..419B}
{Blandford}, R.~D. \& {McKee}, C.~F. 1982, \apj, 255, 419

\bibitem[{{Chelouche} \& {Daniel}(2012)}]{2012ApJ...747...62C}
{Chelouche}, D. \& {Daniel}, E. 2012, \apj, 747, 62

\bibitem[{{Choloniewski}(1981)}]{1981AcA....31..293C}
{Choloniewski}, J. 1981, \actaa, 31, 293

\bibitem[{{Collin} {et~al.}(2006){Collin}, {Kawaguchi}, {Peterson}, \&
  {Vestergaard}}]{2006A&A...456...75C}
{Collin}, S., {Kawaguchi}, T., {Peterson}, B.~M., \& {Vestergaard}, M. 2006,
  \aap, 456, 75

\bibitem[{{Denney} {et~al.}(2009{\natexlab{a}}){Denney}, {Peterson},
  {Dietrich}, {Vestergaard}, \& {Bentz}}]{2009ApJ...692..246D}
{Denney}, K.~D., {Peterson}, B.~M., {Dietrich}, M., {Vestergaard}, M., \&
  {Bentz}, M.~C. 2009{\natexlab{a}}, \apj, 692, 246

\bibitem[{{Denney} {et~al.}(2009{\natexlab{b}}){Denney}, {Peterson}, {Pogge},
  {Adair}, {Atlee}, {Au-Yong}, {Bentz}, {Bird}, {Brokofsky}, {Chisholm},
  {Comins}, {Dietrich}, {Doroshenko}, {Eastman}, {Efimov}, {Ewald}, {Ferbey},
  {Gaskell}, {Hedrick}, {Jackson}, {Klimanov}, {Klimek}, {Kruse},
  {Lad{\'e}route}, {Lamb}, {Leighly}, {Minezaki}, {Nazarov}, {Onken},
  {Petersen}, {Peterson}, {Poindexter}, {Sakata}, {Schlesinger}, {Sergeev},
  {Skolski}, {Stieglitz}, {Tobin}, {Unterborn}, {Vestergaard}, {Watkins},
  {Watson}, \& {Yoshii}}]{2009ApJ...704L..80D}
{Denney}, K.~D., {Peterson}, B.~M., {Pogge}, R.~W., {et~al.}
  2009{\natexlab{b}}, \apjl, 704, L80

\bibitem[{{Dimitrijevi{\'c}} {et~al.}(2007){Dimitrijevi{\'c}}, {Popovi{\'c}},
  {Kova{\v c}evi{\'c}}, {Da{\v c}i{\'c}}, \& {Ili{\'c}}}]{2007MNRAS.374.1181D}
{Dimitrijevi{\'c}}, M.~S., {Popovi{\'c}}, L.~{\v C}., {Kova{\v c}evi{\'c}}, J.,
  {Da{\v c}i{\'c}}, M., \& {Ili{\'c}}, D. 2007, \mnras, 374, 1181

\bibitem[{{Doroshenko} {et~al.}(2008){Doroshenko}, {Sergeev}, \&
  {Pronik}}]{2008ARep...52..167D}
{Doroshenko}, V.~T., {Sergeev}, S.~G., \& {Pronik}, V.~I. 2008, Astronomy
  Reports, 52, 167

\bibitem[{{Edelson} \& {Krolik}(1988)}]{1988ApJ...333..646E}
{Edelson}, R.~A. \& {Krolik}, J.~H. 1988, \apj, 333, 646

\bibitem[{{Gaskell}(2010)}]{2010ASPC..427...68G}
{Gaskell}, C.~M. 2010, Accretion and Ejection in AGN: a Global
  View.~Proceedings of a conference held June 22-26, 2009 in Como,
  Italy.~Edited by Laura Maraschi, Gabriele Ghisellini, Roberto Della Ceca, and
  Fabrizio Tavecchio., p.68, 427, 68

\bibitem[{{Graham} {et~al.}(2011){Graham}, {Onken}, {Athanassoula}, \&
  {Combes}}]{2011MNRAS.412.2211G}
{Graham}, A.~W., {Onken}, C.~A., {Athanassoula}, E., \& {Combes}, F. 2011,
  \mnras, 412, 2211

\bibitem[{{Grier} {et~al.}(2012){Grier}, {Peterson}, {Pogge}, {Denney},
  {Bentz}, {Martini}, {Sergeev}, {Kaspi}, {Minezaki}, {Zu}, {Kochanek},
  {Siverd}, {Shappee}, {Stanek}, {Araya Salvo}, {Beatty}, {Bird}, {Bord},
  {Borman}, {Che}, {Chen}, {Cohen}, {Dietrich}, {Doroshenko}, {Drake},
  {Efimov}, {Free}, {Ginsburg}, {Henderson}, {King}, {Koshida}, {Mogren},
  {Molina}, {Mosquera}, {Nazarov}, {Okhmat}, {Pejcha}, {Rafter}, {Shields},
  {Skowron}, {Szczygiel}, {Valluri}, \& {van Saders}}]{2012arXiv1206.6523G}
{Grier}, C.~J., {Peterson}, B.~M., {Pogge}, R.~W., {et~al.} 2012, ArXiv
  e-prints

\bibitem[{{Haas} {et~al.}(2011){Haas}, {Chini}, {Ramolla}, {Pozo Nu{\~n}ez},
  {Westhues}, {Watermann}, {Hoffmeister}, \& {Murphy}}]{2011A&A...535A..73H}
{Haas}, M., {Chini}, R., {Ramolla}, M., {et~al.} 2011, \aap, 535, A73

\bibitem[{{Isobe} {et~al.}(1990){Isobe}, {Feigelson}, {Akritas}, \&
  {Babu}}]{1990ApJ...364..104I}
{Isobe}, T., {Feigelson}, E.~D., {Akritas}, M.~G., \& {Babu}, G.~J. 1990, \apj,
  364, 104

\bibitem[{{Kaspi} {et~al.}(2005){Kaspi}, {Maoz}, {Netzer}, {Peterson},
  {Vestergaard}, \& {Jannuzi}}]{2005ApJ...629...61K}
{Kaspi}, S., {Maoz}, D., {Netzer}, H., {et~al.} 2005, \apj, 629, 61

\bibitem[{{Kaspi} {et~al.}(2000){Kaspi}, {Smith}, {Netzer}, {Maoz}, {Jannuzi},
  \& {Giveon}}]{2000ApJ...533..631K}
{Kaspi}, S., {Smith}, P.~S., {Netzer}, H., {et~al.} 2000, \apj, 533, 631

\bibitem[{{Koratkar} \& {Gaskell}(1991)}]{1991ApJ...370L..61K}
{Koratkar}, A.~P. \& {Gaskell}, C.~M. 1991, \apjl, 370, L61

\bibitem[{{Landolt}(2009)}]{2009AJ....137.4186L}
{Landolt}, A.~U. 2009, \aj, 137, 4186

\bibitem[{{Lyutyi}(1979)}]{1979SvA....23..518L}
{Lyutyi}, V.~M. 1979, \sovast, 23, 518

\bibitem[{{Netzer}(1990)}]{1990agn..conf...57N}
{Netzer}, H. 1990, in Active Galactic Nuclei, ed. R.~D. {Blandford},
  H.~{Netzer}, L.~{Woltjer}, T.~J.-L. {Courvoisier}, \& M.~{Mayor}, 57--160

\bibitem[{{Onken} {et~al.}(2004){Onken}, {Ferrarese}, {Merritt}, {Peterson},
  {Pogge}, {Vestergaard}, \& {Wandel}}]{2004ApJ...615..645O}
{Onken}, C.~A., {Ferrarese}, L., {Merritt}, D., {et~al.} 2004, \apj, 615, 645

\bibitem[{{Peng} {et~al.}(2002){Peng}, {Ho}, {Impey}, \&
  {Rix}}]{2002AJ....124..266P}
{Peng}, C.~Y., {Ho}, L.~C., {Impey}, C.~D., \& {Rix}, H.-W. 2002, \aj, 124, 266

\bibitem[{{Peterson} {et~al.}(2004){Peterson}, {Ferrarese}, {Gilbert}, {Kaspi},
  {Malkan}, {Maoz}, {Merritt}, {Netzer}, {Onken}, {Pogge}, {Vestergaard}, \&
  {Wandel}}]{2004ApJ...613..682P}
{Peterson}, B.~M., {Ferrarese}, L., {Gilbert}, K.~M., {et~al.} 2004, \apj, 613,
  682

\bibitem[{{Peterson} {et~al.}(1998{\natexlab{a}}){Peterson}, {Wanders},
  {Bertram}, {Hunley}, {Pogge}, \& {Wagner}}]{1998ApJ...501...82P}
{Peterson}, B.~M., {Wanders}, I., {Bertram}, R., {et~al.} 1998{\natexlab{a}},
  \apj, 501, 82

\bibitem[{{Peterson} {et~al.}(1998{\natexlab{b}}){Peterson}, {Wanders},
  {Horne}, {Collier}, {Alexander}, {Kaspi}, \& {Maoz}}]{1998PASP..110..660P}
{Peterson}, B.~M., {Wanders}, I., {Horne}, K., {et~al.} 1998{\natexlab{b}},
  \pasp, 110, 660

\bibitem[{{Sakata} {et~al.}(2010){Sakata}, {Minezaki}, {Yoshii}, {Kobayashi},
  {Koshida}, {Aoki}, {Enya}, {Tomita}, {Suganuma}, {Katsuno Uchimoto}, \&
  {Sugawara}}]{2010ApJ...711..461S}
{Sakata}, Y., {Minezaki}, T., {Yoshii}, Y., {et~al.} 2010, \apj, 711, 461

\bibitem[{{Schlegel} {et~al.}(1998){Schlegel}, {Finkbeiner}, \&
  {Davis}}]{1998ApJ...500..525S}
{Schlegel}, D.~J., {Finkbeiner}, D.~P., \& {Davis}, M. 1998, \apj, 500, 525

\bibitem[{{Storey} \& {Zeippen}(2000)}]{2000MNRAS.312..813S}
{Storey}, P.~J. \& {Zeippen}, C.~J. 2000, \mnras, 312, 813

\bibitem[{{Suganuma} {et~al.}(2006){Suganuma}, {Yoshii}, {Kobayashi},
  {Minezaki}, {Enya}, {Tomita}, {Aoki}, {Koshida}, \&
  {Peterson}}]{2006ApJ...639...46S}
{Suganuma}, M., {Yoshii}, Y., {Kobayashi}, Y., {et~al.} 2006, \apj, 639, 46

\bibitem[{{Vestergaard}(2002)}]{2002ApJ...571..733V}
{Vestergaard}, M. 2002, \apj, 571, 733

\bibitem[{{Watson} {et~al.}(2011){Watson}, {Denney}, {Vestergaard}, \&
  {Davis}}]{2011ApJ...740L..49W}
{Watson}, D., {Denney}, K.~D., {Vestergaard}, M., \& {Davis}, T.~M. 2011,
  \apjl, 740, L49

\bibitem[{{Webb} {et~al.}(1988){Webb}, {Smith}, {Leacock}, {Fitzgibbons},
  {Gombola}, \& {Shepherd}}]{1988AJ.....95..374W}
{Webb}, J.~R., {Smith}, A.~G., {Leacock}, R.~J., {et~al.} 1988, \aj, 95, 374

\bibitem[{{Winkler}(1997)}]{1997MNRAS.292..273W}
{Winkler}, H. 1997, \mnras, 292, 273

\bibitem[{{Winkler} {et~al.}(1992){Winkler}, {Glass}, {van Wyk}, {Marang},
  {Jones}, {Buckley}, \& {Sekiguchi}}]{1992MNRAS.257..659W}
{Winkler}, H., {Glass}, I.~S., {van Wyk}, F., {et~al.} 1992, \mnras, 257, 659

\bibitem[{{Woo} {et~al.}(2007){Woo}, {Treu}, {Malkan}, {Ferry}, \&
  {Misch}}]{2007ApJ...661...60W}
{Woo}, J.-H., {Treu}, T., {Malkan}, M.~A., {Ferry}, M.~A., \& {Misch}, T. 2007,
  \apj, 661, 60

\bibitem[{{Yoshii}(2002)}]{2002ntto.conf..235Y}
{Yoshii}, Y. 2002, in New Trends in Theoretical and Observational Cosmology,
  ed. K.~{Sato} \& T.~{Shiromizu}, 235

\bibitem[{{Yoshii} {et~al.}(2003){Yoshii}, {Kobayashi}, \&
  {Minezaki}}]{2003AAS...202.3803Y}
{Yoshii}, Y., {Kobayashi}, Y., \& {Minezaki}, T. 2003, in Bulletin of the
  American Astronomical Society, Vol.~35, American Astronomical Society Meeting
  Abstracts \#202, 752

\end{thebibliography}

\onllongtab{5}{
  \begin{longtable}{ccccc}
    \caption{$B$, $V$, $OIII$ and H$\beta$ Fluxes in mJy.}\\
    \label{table5}
    JD-2,450,000 & $F_{B}$ & $F_{V}$ & $F_{OIII}$ & $F_{H\beta}$ \\
    \endfirsthead
    \caption{continued.} \\
    JD-2,450,000 & $F_{B}$ & $F_{V}$ & $F_{OIII}$ & $F_{H\beta}$ \\
    \endhead
    55110.250 & $1.870\pm  0.032$ & $3.443\pm  0.033$ &  $5.896\pm  0.018$ & $4.174\pm  0.038$ \\
    55111.250 & $-    \pm  -    $ & $3.487\pm  0.053$ &  $6.051\pm  0.060$ & $4.308\pm  0.081$ \\
    55112.250 & $1.978\pm  0.038$ & $3.524\pm  0.051$ &  $6.009\pm  0.065$ & $4.247\pm  0.083$ \\
    55113.238 & $1.985\pm  0.034$ & $3.555\pm  0.062$ &  $6.047\pm  0.069$ & $4.269\pm  0.093$ \\
    55114.230 & $2.074\pm  0.010$ & $3.606\pm  0.027$ &  $6.216\pm  0.065$ & $4.413\pm  0.070$ \\
    55115.230 & $2.104\pm  0.012$ & $3.653\pm  0.055$ &  $6.276\pm  0.054$ & $4.450\pm  0.077$ \\
    55121.289 & $2.165\pm  0.024$ & $3.700\pm  0.031$ &  $6.108\pm  0.043$ & $4.258\pm  0.053$ \\
    55123.211 & $2.208\pm  0.035$ & $3.721\pm  0.036$ &  $6.240\pm  0.055$ & $4.379\pm  0.066$ \\
    55124.219 & $2.265\pm  0.019$ & $3.769\pm  0.026$ &  $6.299\pm  0.058$ & $4.414\pm  0.064$ \\
    55125.219 & $2.242\pm  0.014$ & $3.789\pm  0.050$ &  $6.283\pm  0.080$ & $4.388\pm  0.094$ \\
    55126.219 & $2.257\pm  0.010$ & $3.790\pm  0.042$ &  $6.293\pm  0.083$ & $4.398\pm  0.093$ \\
    55127.270 & $2.256\pm  0.011$ & $3.808\pm  0.046$ &  $6.456\pm  0.096$ & $4.552\pm  0.107$ \\
    55130.172 & $2.235\pm  0.015$ & $3.805\pm  0.043$ &  $6.475\pm  0.091$ & $4.573\pm  0.100$ \\
    55131.180 & $2.216\pm  0.022$ & $3.786\pm  0.030$ &  $6.523\pm  0.088$ & $4.629\pm  0.094$ \\
    55132.172 & $2.245\pm  0.020$ & $3.798\pm  0.084$ &  $6.569\pm  0.073$ & $4.670\pm  0.112$ \\
    55133.211 & $2.280\pm  0.008$ & $3.824\pm  0.018$ &  $6.558\pm  0.116$ & $4.646\pm  0.118$ \\
    55134.180 & $2.311\pm  0.014$ & $3.843\pm  0.024$ &  $6.487\pm  0.123$ & $4.565\pm  0.126$ \\
    55135.180 & $2.389\pm  0.006$ & $3.911\pm  0.018$ &  $6.510\pm  0.176$ & $4.555\pm  0.177$ \\
    55136.172 & $2.438\pm  0.023$ & $3.947\pm  0.029$ &  $6.471\pm  0.211$ & $4.498\pm  0.213$ \\
    55137.191 & $2.523\pm  0.027$ & $3.980\pm  0.065$ &  $6.489\pm  0.245$ & $4.498\pm  0.254$ \\
    55138.172 & $2.539\pm  0.058$ & $4.000\pm  0.111$ &  $6.595\pm  0.203$ & $4.595\pm  0.232$ \\
    55139.172 & $2.540\pm  0.094$ & $4.005\pm  0.053$ &  $6.696\pm  0.212$ & $4.693\pm  0.218$ \\
    55140.191 & $2.513\pm  0.034$ & $4.004\pm  0.038$ &  $6.750\pm  0.167$ & $4.748\pm  0.171$ \\
    55141.191 & $2.492\pm  0.024$ & $4.028\pm  0.022$ &  $6.786\pm  0.149$ & $4.771\pm  0.150$ \\
    55144.148 & $2.415\pm  0.026$ & $4.027\pm  0.038$ &  $6.788\pm  0.124$ & $4.774\pm  0.130$ \\
    55146.191 & $2.400\pm  0.029$ & $4.055\pm  0.042$ &  $6.806\pm  0.129$ & $4.778\pm  0.135$ \\
    55147.160 & $2.394\pm  0.020$ & $4.056\pm  0.024$ &  $6.930\pm  0.135$ & $4.902\pm  0.137$ \\
    55148.180 & $2.370\pm  0.036$ & $4.054\pm  0.085$ &  $6.942\pm  0.132$ & $4.915\pm  0.157$ \\
    55149.191 & $2.375\pm  0.031$ & $4.042\pm  0.045$ &  $6.912\pm  0.131$ & $4.891\pm  0.139$ \\
    55151.180 & $2.386\pm  0.056$ & $4.028\pm  0.021$ &  $6.994\pm  0.120$ & $4.980\pm  0.122$ \\
    55153.160 & $2.385\pm  0.042$ & $4.041\pm  0.032$ &  $6.935\pm  0.129$ & $4.914\pm  0.133$ \\
    55154.180 & $2.416\pm  0.034$ & $4.064\pm  0.028$ &  $6.820\pm  0.131$ & $4.788\pm  0.134$ \\
    55156.141 & $2.449\pm  0.043$ & $4.069\pm  0.059$ &  $6.892\pm  0.159$ & $4.857\pm  0.170$ \\
    55157.180 & $2.473\pm  0.068$ & $4.095\pm  0.045$ &  $7.043\pm  0.164$ & $4.995\pm  0.170$ \\
    55158.180 & $2.488\pm  0.052$ & $-    \pm  -    $ &  $6.962\pm  0.179$ & $4.915\pm  0.183$ \\
    55159.160 & $2.511\pm  0.024$ & $4.111\pm  0.038$ &  $7.013\pm  0.167$ & $4.957\pm  0.176$ \\
    55160.230 & $2.516\pm  0.015$ & $4.116\pm  0.055$ &  $7.063\pm  0.134$ & $5.004\pm  0.138$ \\
    55161.160 & $2.510\pm  0.029$ & $4.061\pm  0.034$ &  $7.007\pm  0.111$ & $4.977\pm  0.112$ \\
    55162.148 & $2.516\pm  0.020$ & $4.073\pm  0.012$ &  $6.937\pm  0.095$ & $4.900\pm  0.155$ \\
    55163.211 & $2.497\pm  0.030$ & $4.070\pm  0.122$ &  $7.016\pm  0.097$ & $4.980\pm  0.103$ \\
    55164.160 & $2.512\pm  0.039$ & $4.110\pm  0.036$ &  $7.108\pm  0.094$ & $5.053\pm  0.109$ \\
    55165.172 & $2.505\pm  0.023$ & $4.090\pm  0.055$ &  $7.116\pm  0.110$ & $5.071\pm  0.116$ \\
    55166.148 & $2.552\pm  0.101$ & $-    \pm  -    $ &  $7.158\pm  0.100$ & $5.112\pm  0.107$ \\
    55167.141 & $2.546\pm  0.018$ & $4.132\pm  0.038$ &  $7.226\pm  0.090$ & $5.159\pm  0.098$ \\
    55168.180 & $2.581\pm  0.025$ & $4.090\pm  0.038$ &  $7.281\pm  0.086$ & $5.235\pm  0.097$ \\
    55173.180 & $2.564\pm  0.067$ & $4.109\pm  0.036$ &  $7.265\pm  0.083$ & $5.211\pm  0.092$ \\
    55174.262 & $2.477\pm  0.144$ & $4.051\pm  0.045$ &  $7.188\pm  0.076$ & $5.163\pm  0.088$ \\
    55175.172 & $-    \pm  -    $ & $-    \pm  -    $ &  $7.235\pm  0.068$ & $5.210\pm  0.070$ \\
    55176.219 & $-    \pm  -    $ & $-    \pm  -    $ &  $7.233\pm  0.078$ & $5.208\pm  0.101$ \\
    55177.180 & $2.376\pm  0.055$ & $4.051\pm  0.038$ &  $7.235\pm  0.073$ & $5.209\pm  0.103$ \\
    55179.199 & $2.384\pm  0.029$ & $4.053\pm  0.044$ &  $7.281\pm  0.078$ & $5.254\pm  0.091$ \\
    55180.180 & $2.266\pm  0.022$ & $3.979\pm  0.012$ &  $7.199\pm  0.087$ & $5.210\pm  0.094$ \\
    55181.180 & $2.147\pm  0.015$ & $3.857\pm  0.064$ &  $7.028\pm  0.091$ & $5.099\pm  0.128$ \\
    55200.160 & $2.258\pm  0.019$ & $3.910\pm  0.072$ &  $6.988\pm  0.090$ & $5.033\pm  0.092$ \\
    55201.141 & $2.226\pm  0.017$ & $3.815\pm  0.046$ &  $6.880\pm  0.112$ & $4.972\pm  0.118$ \\
    55202.160 & $2.059\pm  0.213$ & $3.693\pm  0.036$ &  $6.602\pm  0.130$ & $4.755\pm  0.130$ \\
    55203.141 & $2.045\pm  0.056$ & $3.677\pm  0.089$ &  $6.586\pm  0.135$ & $4.748\pm  0.137$ \\
    55204.172 & $2.065\pm  0.034$ & $3.639\pm  0.017$ &  $6.563\pm  0.148$ & $4.744\pm  0.148$ \\
    55205.129 & $1.912\pm  0.019$ & $3.494\pm  0.035$ &  $6.420\pm  0.149$ & $4.673\pm  0.156$ \\
    55206.148 & $1.858\pm  0.022$ & $3.463\pm  0.008$ &  $6.340\pm  0.134$ & $4.608\pm  0.139$ \\
    55207.141 & $1.869\pm  0.020$ & $3.450\pm  0.025$ &  $6.342\pm  0.135$ & $4.617\pm  0.136$ \\
    55208.141 & $1.860\pm  0.011$ & $3.425\pm  0.007$ &  $6.315\pm  0.116$ & $4.602\pm  0.119$ \\
    55209.148 & $1.834\pm  0.015$ & $3.421\pm  0.043$ &  $6.299\pm  0.100$ & $4.588\pm  0.115$ \\
    55210.148 & $1.819\pm  0.015$ & $3.383\pm  0.037$ &  $6.250\pm  0.101$ & $4.559\pm  0.105$ \\
    55212.129 & $1.822\pm  0.017$ & $3.357\pm  0.015$ &  $6.241\pm  0.081$ & $4.562\pm  0.083$ \\
    55213.141 & $1.847\pm  0.011$ & $3.347\pm  0.025$ &  $6.239\pm  0.072$ & $4.565\pm  0.077$ \\
    55214.129 & $1.849\pm  0.010$ & $3.361\pm  0.055$ &  $6.222\pm  0.072$ & $4.541\pm  0.076$ \\
    55215.129 & $1.856\pm  0.018$ & $3.383\pm  0.029$ &  $6.179\pm  0.068$ & $4.487\pm  0.078$ \\
    55216.141 & $1.880\pm  0.008$ & $3.409\pm  0.021$ &  $6.230\pm  0.066$ & $4.526\pm  0.070$ \\
    55217.109 & $1.908\pm  0.008$ & $3.423\pm  0.027$ &  $6.162\pm  0.062$ & $4.450\pm  0.090$ \\
    55218.109 & $1.929\pm  0.007$ & $3.436\pm  0.023$ &  $6.190\pm  0.058$ & $4.472\pm  0.058$ \\
    55219.129 & $1.941\pm  0.020$ & $3.468\pm  0.037$ &  $6.242\pm  0.066$ & $4.508\pm  0.080$ \\
    55220.109 & $1.945\pm  0.012$ & $3.434\pm  0.023$ &  $6.327\pm  0.084$ & $4.610\pm  0.090$ \\
    55221.121 & $1.974\pm  0.026$ & $3.452\pm  0.065$ &  $6.381\pm  0.104$ & $4.655\pm  0.108$ \\
    55222.121 & $2.005\pm  0.035$ & $3.474\pm  0.009$ &  $6.414\pm  0.148$ & $4.677\pm  0.148$ \\
    55223.090 & $1.995\pm  0.035$ & $3.486\pm  0.045$ &  $6.414\pm  0.154$ & $4.671\pm  0.154$ \\
    55224.102 & $1.975\pm  0.049$ & $3.431\pm  0.033$ &  $6.375\pm  0.139$ & $4.660\pm  0.144$ \\
    55225.109 & $1.956\pm  0.018$ & $3.444\pm  0.027$ &  $6.259\pm  0.125$ & $4.537\pm  0.125$ \\
    55226.090 & $1.917\pm  0.030$ & $3.418\pm  0.006$ &  $6.152\pm  0.107$ & $4.443\pm  0.107$ \\
    55228.109 & $1.890\pm  0.006$ & $3.395\pm  0.004$ &  $6.113\pm  0.071$ & $4.415\pm  0.078$ \\
    55230.078 & $1.880\pm  0.032$ & $3.397\pm  0.036$ &  $6.062\pm  0.063$ & $4.363\pm  0.072$ \\
    55231.090 & $1.897\pm  0.019$ & $3.418\pm  0.010$ &  $6.005\pm  0.072$ & $4.296\pm  0.093$ \\
    55232.078 & $1.911\pm  0.017$ & $3.431\pm  0.009$ &  $5.954\pm  0.084$ & $4.238\pm  0.086$ \\
    55233.078 & $1.944\pm  0.151$ & $3.448\pm  0.032$ &  $5.960\pm  0.092$ & $4.236\pm  0.093$ \\
    55234.121 & $-    \pm  -    $ & $-    \pm  -    $ &  $5.928\pm  0.094$ & $4.204\pm  0.101$ \\
    55236.070 & $1.966\pm  0.018$ & $3.484\pm  0.034$ &  $5.957\pm  0.103$ & $4.215\pm  0.120$ \\
    55238.070 & $2.019\pm  0.032$ & $3.494\pm  0.059$ &  $6.027\pm  0.089$ & $4.280\pm  0.090$ \\
    55240.070 & $2.061\pm  0.028$ & $3.544\pm  0.018$ &  $6.136\pm  0.101$ & $4.364\pm  0.101$ \\
    55241.051 & $2.105\pm  0.044$ & $3.585\pm  0.017$ &  $6.174\pm  0.106$ & $4.381\pm  0.106$ \\
    55242.051 & $2.114\pm  0.021$ & $3.622\pm  0.038$ &  $6.281\pm  0.102$ & $4.470\pm  0.102$ \\
    55243.059 & $2.131\pm  0.018$ & $3.635\pm  0.061$ &  $6.261\pm  0.099$ & $4.444\pm  0.099$ \\
    55244.039 & $2.168\pm  0.011$ & $3.654\pm  0.015$ &  $6.296\pm  0.116$ & $4.469\pm  0.116$ \\
    55245.059 & $2.159\pm  0.020$ & $3.643\pm  0.014$ &  $6.355\pm  0.099$ & $4.533\pm  0.099$ \\
    55246.059 & $-    \pm  -    $ & $-    \pm  -    $ &  $6.394\pm  0.092$ & $4.572\pm  0.092$ \\
    55250.051 & $-    \pm  -    $ & $-    \pm  -    $ &  $6.343\pm  0.121$ & $4.523\pm  0.121$ \\
    55251.051 & $2.332\pm  0.013$ & $-    \pm  -    $ &  $6.282\pm  0.137$ & $4.462\pm  0.137$ \\
    55252.051 & $-    \pm  -    $ & $3.640\pm  0.017$ &  $6.273\pm  0.123$ & $4.453\pm  0.123$ \\
    55254.039 & $-    \pm  -    $ & $3.608\pm  0.013$ &  $6.203\pm  0.122$ & $4.399\pm  0.122$ \\
    55258.051 & $-    \pm  -    $ & $-    \pm  -    $ &  $6.130\pm  0.121$ & $4.326\pm  0.121$ \\
    55261.051 & $-    \pm  -    $ & $-    \pm  -    $ &  $6.148\pm  0.094$ & $4.343\pm  0.094$ \\
    55262.039 & $-    \pm  -    $ & $-    \pm  -    $ &  $6.067\pm  0.097$ & $4.263\pm  0.097$ \\
    55263.039 & $-    \pm  -    $ & $-    \pm  -    $ &  $6.232\pm  0.063$ & $4.428\pm  0.063$ \\
  \end{longtable}
}

\end{document}